\documentclass[useAMS,usenatbib]{mn2e}

\usepackage{verbatim,graphicx,epsfig,dcolumn}
\newcolumntype{.}{D{.}{.}{4}}
\newcolumntype{,}{D{.}{.}{2}}
\newcolumntype{;}{D{.}{.}{1}}
\newcommand{\nodata}{$\cdot\cdot\cdot$}
\newcommand{\lesssim}{{\lower-1.2pt\vbox{\hbox{\rlap{$<$}\lower5pt\vbox{\hbox{$\sim$}}}}}}
\newcommand{\gtrsim}{{\lower-1.2pt\vbox{\hbox{\rlap{$>$}\lower5pt\vbox{\hbox{$\sim$}}}}}}
\title[Carbon enrichment in the Sgr dSph]{Carbon enrichment of the evolved stars in the Sagittarius dwarf spheroidal}
\author[I. McDonald et al.]{I.~McDonald$^{1}$\thanks{E-mail:
mcdonald@jb.man.ac.uk}, J.~R.~White$^{1}$, A.~A.~Zijlstra$^{1}$, L.~Guzman Ramirez$^{1}$, C.~Szyszka$^{1,2}$, \newauthor
J.~Th.~van Loon$^{3}$, E.~Lagadec$^{4}$, O.~C.~Jones$^{1}$\\
$^{1}$Jodrell Bank Centre for Astrophysics, Alan Turing Building, Manchester, M13 9PL, UK\\
$^{2}$Institute for Astro- and Particle Physics, Technikerstr.\ 25/8, 6020 Innsbruck, Austria\\
$^{3}$Lennard-Jones Laboratories, Keele University, ST5 5BG, UK\\
$^{4}$European Southern Observatory, Karl Schwarzschildstrasse 2, Garching 85748, Germany}

\begin{document}

\date{Accepted 9999 December 32. Received 9999 December 32; in original form 9999 December 32}

\pagerange{\pageref{firstpage}--\pageref{lastpage}} \pubyear{9999}

\maketitle

\label{firstpage}

\begin{abstract}
We present spectra of 1142 colour-selected stars in the direction of the Sagittarius Dwarf Spheroidal (Sgr dSph) galaxy, of which 1058 were taken with VLT/FLAMES multi-object spectrograph and 84 were taken with the SAAO Radcliffe 1.9-m telescope grating spectrograph. Spectroscopic membership is confirmed (at $>$99\% confidence) for 592 stars on the basis of their radial velocity, and spectral types are given. Very slow rotation is marginally detected around the galaxy's major axis. We identify five S stars and 23 carbon stars, of which all but four carbon stars are newly-determined and all but one (PQ Sgr) are likely Sgr dSph members. We examine the onset of carbon-richness in this metal-poor galaxy in the context of stellar models. We compare the stellar death rate (one star per 1000--1700 years) to known planetary nebula dynamical ages and find that the bulk population produce the observed (carbon-rich) planetary nebulae. We compute average lifetimes of S and carbon stars as 60--250 and 130--500 kyr, compared to a total thermal-pulsing asymptotic giant branch lifetime of 530--1330 kyr. We conclude by discussing the return of carbon-rich material to the ISM.
\end{abstract}



\section{Introduction}
\label{IntroSect}

The mass return from stars to the interstellar medium (ISM) is a powerful driver of chemical change in a galaxy. The abundance and amount of material returned to the ISM determines the chemical make-up of successive generations of stars and planets. A large fraction of the metals returned to the ISM are ejected by highly-evolved asymptotic giant branch (AGB) stars (e.g.\ \citealt{Gehrz89,ZGT08}) which lose mass via pulsation-enhanced, dust-driven winds (e.g.\ \citealt{Woitke06a}). Dusty material returned by oxygen-rich stars include silicates, primarily pyroxenes and olivines (Mg$_{1-x}$Fe$_{x}$SiO$_3$, Mg$_{2-2x}$Fe$_{2x}$SiO$_4$), oxides such as Al$_2$O$_3$, and potentially metals and metal alloys (e.g.\ \citealt{SKGP03,vLMO+06,MvLD+09,TMB+09,MSZ+10,MvLS+11,NTI+12}). Carbon stars return primarily amorphous carbon, silicon carbide, and other carbon-rich species \citep{TC74,MHB00,LZS+07}. Carbon-rich dust species tend to have significantly higher opacities than oxygen-rich species, meaning that stellar outflows can be driven more effectively \citep{Woitke06b}. It has been proposed that the `superwind' (the final, high-mass-loss-rate phase of the AGB wind that ejects the remainder of the star's atmosphere and ends its nuclear-burning life; \citealt{vLGdK+99,Willson00,FG06}) may sometimes be triggered by the increase in C/O ratio \citep{WlBJ+00,LZ08}.

Giant stars undergo dredge up of fusion-processed material from their cores at key stages in their post-main-sequence evolution (e.g.\ \citealt{SL85,Mowlavi99,Herwig04}). Carbon is an important element in the dredge-up process, particularly during third dredge-up, which occurs on the thermally-pulsing AGB (TP-AGB) \citep{DTKL03}. It is over-abundant in the dredged material, which can result in an increase in the carbon-to-oxygen (C/O) ratio from a main-sequence value of C/O $\approx 0.4$ to C/O $>$ 1. Oxygen and carbon are among the most abundant metals, and CO is one of the first molecules to form. This binds away free carbon and oxygen, leaving the more abundant element to dominate the stellar chemistry. When C/O exceeds unity (by number), the star transforms from an oxygen-rich star to a carbon-rich star. The length of this carbon-rich period (which lasts for the remainder of the star's nuclear-burning life) and the mass loss that occurs during it, directly affects how much carbon is returned to the interstellar medium.

Of particular importance for the historical enrichment of the ISM, in our own and other galaxies, is the variation of the onset and length of this carbon-rich phase with metallicity and initial stellar mass. Only some stars become carbon-rich: low-mass stars do not dredge up sufficient carbon to become carbon-rich, while high-mass stars undergo hot-bottom burning via the CNO cycle, which reduces the carbon abundance in the dredged-up layers, limiting surface carbon enrichment \citep{LFCW96}. Exactly where these limits lie depends on the complex parameters of stellar mass loss and dredge-up \citep{VM10}. For stars of (initially) solar abundance, the bounds for becoming carbon rich lie at initial masses of approximately 2 and 4 M$_\odot$ \citep{MG07}. Lower-metallicity stars, which have lower initial oxygen abundances, require less carbon to be dredged up to become carbon-rich. The mass range of low-metallicity carbon stars is therefore expected to be much larger (e.g.\ \citealt{MZvL+05,ZMW+06,GM07}).

The onset of carbon-richness at different metallicities and masses can be probed by examining stellar populations with different metallicities and ages. Among the most interesting is the Sagittarius Dwarf Spheroidal (Sgr dSph) or Elliptical Galaxy (Sag DEG; \citealt{IGI94,IGI95,IWG+97}): this metal-poor galaxy lies at a relatively-close distance of 25 kpc \citep{MKS+95,MBFP04,KC09} and is being tidally disrupted by the Milky Way \citep{TI98,MSWO03,LJM05}. The galaxy contains several identified populations. While the delineation between these populations is indistinct, at least at present, they have been broadly summarised by \citet{SDM+07} into five separate groups:
\begin{enumerate}
\item a very metal-poor population with [Fe/H] = --1.7 dex, [$\alpha$/Fe] $\approx$ +0.2 dex, and an age of $t \approx 13$ Gyr, which derives from (and is mostly contained within) the globular cluster M54;
\item a sizable, metal-poor population with [Fe/H] $\approx$ --1.2 dex, $t$ = 10--12 Gyr, most prevalent in the outskirts (see also \citealt{LS00});
\item a bulk population with a spread of metallicity between [Fe/H] = --0.7 to --0.4 dex, [$\alpha$/Fe] $\approx$ --0.2 dex, and ages between $t$ = 4 -- 8 Gyr, which dominates in the galaxy's core (see also \citealt{BCF+06});
\item a small, higher-metallicity population ([Fe/H] $\approx$ --0.4 to --0.1, [$\alpha$/Fe] $\approx$ +0.2 dex, $t$ = 2 -- 3 Gyr; \citealt{SL95,LS00,ZGW+06,SDM+07}); and
\item a probable, very small, metal-rich population ([Fe/H] $\approx$ +0.5, [$\alpha$/Fe] $\approx$ 0 dex, $t < 1$ Gyr, newly discovered by \citealt{SDM+07}).
\end{enumerate}
Several globular clusters may also be historically associated with the galaxy and its tidal stream\footnote{Though authors differ on which clusters are associated, the following are generally accepted: M54, Terzan 7, Terzan 8, Arp 2, Palomar 2 (still uncertain), Palomar 12, Whiting 1 (\citealt{DCA95,LM10} and references therein).}; these have metallicities between [Fe/H] = --2.16 (Terzan 8) and --0.32 (Terzan 7) \citep{Harris96}. The best-studied of these is M54, which lies at the projected centre of the galaxy.

The Sgr dSph's most-evolved AGB stars should be close to the lower mass limit for carbon-richness. Previous works have noted several carbon stars in the galaxy, and linked them to the metal-rich population \citep{LZ08,LZS+09}. In this work, we examine spectra of a significant fraction of the galaxy's giant stars, in order to confirm their membership of the galaxy and determine the point at which stars become carbon rich.


\section{Observations}

\begin{table*}
 \centering
 \begin{minipage}{160mm}
  \caption{Details of observations taken.}
\label{ObsTable}
  \begin{tabular}{@{}l@{\qquad}c@{}c@{}c@{}l@{\qquad}c@{}c@{}r@{}r@{}}
  \hline\hline
   \multicolumn{1}{c}{Telescope}	& \multicolumn{1}{c}{Field}	& \multicolumn{2}{c}{Centre (J2000.0)}	& \multicolumn{1}{c}{Observation}	& \multicolumn{1}{c}{Instrument}	& \multicolumn{1}{c}{Coverage}	& \multicolumn{1}{c}{Resolution} & \multicolumn{1}{c}{Number of}\\
   \  & \  & \multicolumn{1}{c}{RA} & \multicolumn{1}{c}{Dec} & \multicolumn{1}{c}{date}		& \  & \multicolumn{1}{c}{(nm)}	& \multicolumn{1}{c}{($\lambda$/$\Delta\lambda$)} & \multicolumn{1}{c}{objects}\\
 \hline
VLT & \ & \ & \ & \ & \  & \ & \ & \ \\
\  & 1 & 18 56 24 & --30 31 48 & 2010-06-24 & GIRAFFE & 574.1--652.4 &  7\,400 & 1015\\
\  & 2 & 18 56 27 & --31 22 47 & 2010-09-20 & \       & 643.8--718.4 &  8\,600 & \ \\
\  & 3 & 18 59 39 & --31 01 10 & 2010-09-20 & \       & 638.3--662.6 & 28\,800 & \ \\
\  & 4 & 18 54 00 & --30 44 59 & 2010-09-22 & \       & 693.7--725.0 & 23\,900 & \ \\
\  & 5 & 18 53 24 & --30 20 59 & 2010-09-23 & UVES    & 476.0--684.0 & 47\,000 & 43\\
\  & 6 & 18 58 12 & --30 42 00 & 2010-09-24 & \       & \  & \  & \ \\
\  & 7 & 18 55 12 & --30 12 00 & 2010-09-25 & \       & \  & \  & \ \\
\  & 8 & 18 48 29 & --30 12 02 & 2010-09-26 & \       & \  & \  & \ \\
\  & 9 & 18 52 12 & --29 59 58 & 2010-09-27 & \       & \  & \  & \ \\
SAAO-1.9m \\
\  & 1 & \nodata & \nodata & 2009-07-04 to & SpCCD & 400--730 & 1\,750--3\,170 & 28\\
\  & \ & \        & \        & 2009-07-08    & \     & \  & \  & \ \\
\  & 2 & \nodata & \nodata & 2010-07-28 to & SpCCD & 642--722 & 15\,000--16\,800 & 262\\
\  & \ & \        & \        & 2010-08-04    & \     & \  & \  & \ \\
\hline
\end{tabular}
\end{minipage}
\end{table*}

\begin{figure}
 \resizebox{\hsize}{!}{\includegraphics[angle=270]{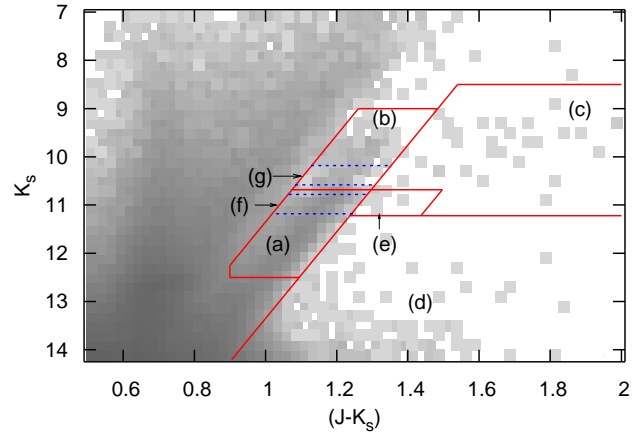}}
 \caption{2MASS colour--magnitude diagram of the Sgr dSph core, showing the colour cuts used to isolate the Sgr dSph RGB (region (a)) and AGB (region (b)). Region (c) is inhabited by dusty AGB stars. Region (d) contains foreground dwarf stars and background galaxies. Region (e) contains a population of unknown providence, which is discussed in Section \protect\ref{EvolRateSect}. The dashed blue lines show the regions (f,g) used to calculate RGB and AGB star counts.}
 \label{CutsFig}
\end{figure}

\begin{figure}
 \resizebox{\hsize}{!}{\includegraphics[angle=270]{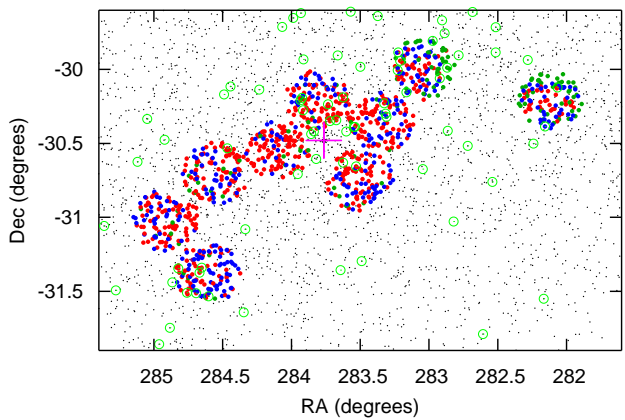}}
 \caption{Spatial distribution of the observed VLT/FLAMES fields, against stars from the 2MASS catalogue with ($J-K_{\rm s}$) colours matching the Sgr dSph giant branch. The open circles show the SAAO targets. The large magenta cross marks the globular cluster M54, the galaxy's nominal centre. Red points denote radial velocity members, blue points denote radial velocity non-members, green points denote stars where a radial velocity was not accurately determined.}
 \label{SpatialFig}
\end{figure}

\begin{figure}
 \resizebox{\hsize}{!}{\includegraphics[angle=270]{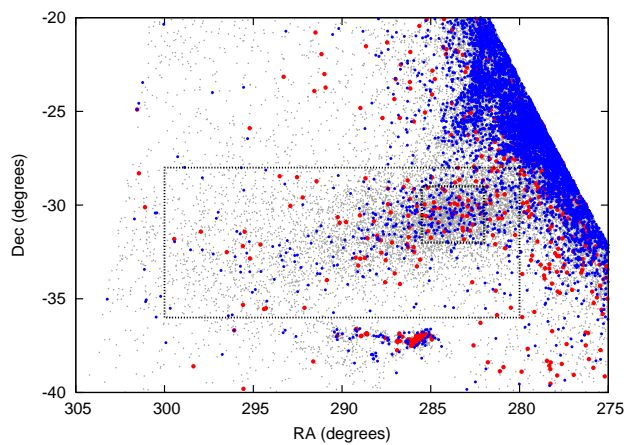}}
 \caption{Location of colour-selected RGB stars (grey dots, region (a) in Figure \protect\ref{CutsFig}). Overlaid as small blue points are (near-) naked AGB stars, selected from region (b) of Figure \protect\ref{CutsFig}. Overlaid as larger red points are dusty AGB stars, selected from regions (c) and (e) of Figure \protect\ref{CutsFig}. The blue, contaminating population to the top right is the Galactic Disc, the group at RA = 285$^\circ$ to 290$^\circ$, Dec = --37$^\circ$ is the Galactic star-forming region associated with NGC 6727. The black lines mark the regions we use to define the galaxy's ``core'' and ``main body'' within the text.}
 \label{AGBFig}
\end{figure}

The observed stars were sampled from the 2MASS catalogue \citep{CSvD+03}. The 2MASS $(J-K_{\rm s})$ vs.\ $K_s$ colour--magnitude diagram (CMD) for the 3$^\circ$ around M54 (which we take to be the centre of the Sgr dSph) shows a distinct feature associable with the Sgr dSph giant branch (\citealt{MSWO03}; see also Figure \ref{CutsFig}). Stars represented by this feature can be separated from the foreground Galactic population using a colour cut of $K_s > 20.35 - 9 (J-K_{\rm s})$ mag (left-most diagonal line, Figure \ref{CutsFig}), which works tolerably well for stars of $K_s < 12$ mag. Priority was given to stars above the RGB tip ($K_s \approx 10.7$ mag).

The limits of allowed fibre positions and field of view of the FLAMES spectrograph meant that these observations contained some stars above the colour cut. These are likely Galactic stars, which are useful for determining the Galactic velocity spread in the region, but these stars may also include some unusual objects from the Sgr dSph. We show the spatial location of our observations within the Sgr dSph in Figure \ref{SpatialFig}, noting that they are very much concentrated in the galaxy's core (Figures \ref{AGBFig}).

The bulk of our observed spectra were obtained using the ESO VLT/FLAMES+MEDUSA multi-fibre spectrograph in GIRAFFE+UVES mode. These cover nine fields of the Sgr dSph core, each taken in four settings (L714.2, L682.2, H651.5B and H710.5)\footnote{Full data on each setting is available in the FLAMES manual: http://www.eso.org/sci/facilities/paranal/instruments/flames/}. Simultaneous spectra were taken using the UVES instrument using the red arm, cross-disperser CD3 and a central wavelength of 580 nm. The higher resolution of UVES allowed us to focus on objects near the AGB tip. We note, however, that the \'echelle format of UVES and the faint nature of many of our targets means that the spectra suffer from substantial distortion where spectral orders meet, limiting our ability to measure the shape of broad molecular features.

The large size of the Sgr dSph compared to the FLAMES field of view (Figures \ref{AGBFig}) prompted observations of additional stars using the grating spectrograph on the SAAO Radcliffe 1.9-m telescope over two runs. Observations were taken during variable conditions, however, leading to a variable signal-to-noise in the final data. TiO or CN bands could be discerned in 69 objects.

These observations are summarised in Table \ref{ObsTable}.

The VLT data were extracted from the standard pipeline reduction provided by ESO. As the GIRAFFE data consist of single exposures in each setup, it was not possible to mitigate against cosmic rays. The data were sky subtracted, but not corrected for telluric absorption as this is not required for our current analysis. The simultaneous UVES data consist of four exposures per setup. The four resulting spectra were median combined to remove cosmic rays.

The SAAO data were reduced following the normal pattern of bias and dark subtraction. Flatfielding was not performed due to the poor quality of flatfields obtained during the observing run. Differences in the responses of adjacent pixels in the detector are small compared to the noise in the final spectra; while lack of flatfielding means we cannot determine the overall shape of the spectra, it should not significantly affect the accuracy of continuum-divided spectra.

Literature data for 324 objects were also sourced from \citet{IGI95,WMIF99}\footnote{We do not include Whitelock et al.'s WMIF 9 and 16--27 as, although they are radial velocity members of the Sgr dSph, they are out of our spatial region of analysis. Additionally, WMIF 13 cannot be identified in the 2MASS catalogue.}; \citet{BHM+00,BSM+04,LMC06,MvL07,LZS+09,GSZ+10,CBG+10}; the General Catalogue of Galactic Carbon Stars \citep{ABD+01}; the General Catalogue of Variable Stars \citep{SDZ+06}; and the Radial Velocity Experiment \citep{SWS+11}. These sources contain a variety of radial velocity and spectral typing data which we use in later figures.


\section{Analysis}

\subsection{Spectral typing}
\label{SpTSect}

\begin{figure*}
 \resizebox{\hsize}{!}{\includegraphics[angle=270]{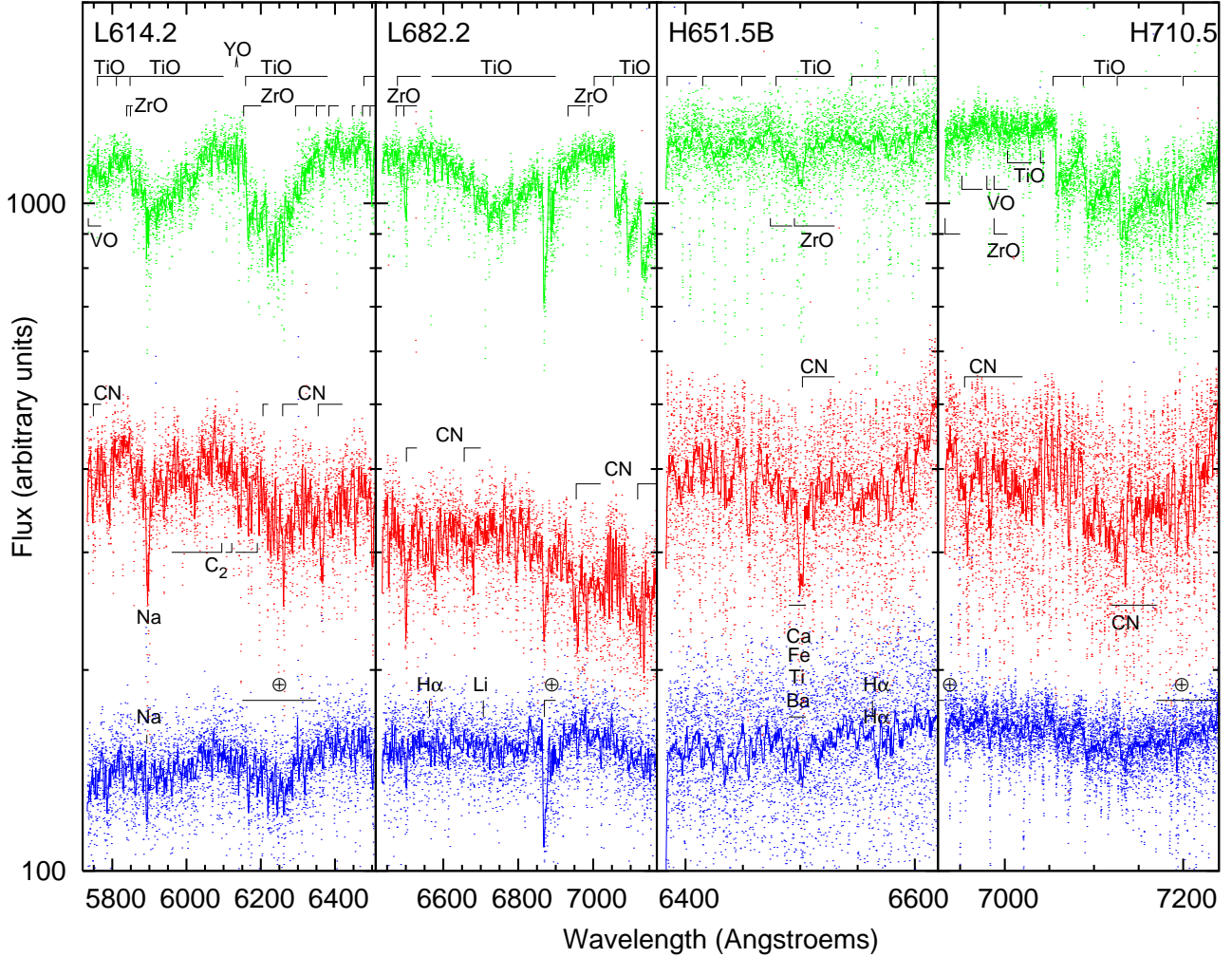}}
 \caption{Examples of typical spectra, selected from stars just below the RGB tip (top: M-type \#105, middle: C-type \#1015, bottom: K-type \#949), in the four observed filter settings. Notable atomic and molecular lines are shown.}
 \label{SpecFig}
\end{figure*}

\begin{table*}
 \centering
 \begin{minipage}{160mm}
  \caption{Spectral types and radial velocities of targets observed with the VLT. Stars with IDs $\geq$1016 were observed with UVES, the remainder with GIRAFFE. The right-hand five columns denote the median heliocentric radial velocity determined from the spectra, the standard deviation of radial velocities determined for that object (which we take as the error), the probability of membership of the Sgr dSph, and the TiO and ZrO band strengths as described in the text. A full table is available in the online version.}
\label{VLTTargets}
  \begin{tabular}{lllcrrrrrcc}
  \hline\hline
   \multicolumn{1}{c}{ID}	& \multicolumn{1}{c}{RA}	& \multicolumn{1}{c}{Dec}	& \multicolumn{1}{c}{Spectral}	& \multicolumn{2}{c}{2MASS}	& \multicolumn{1}{c}{$v_{\rm helio}$}	& \multicolumn{1}{c}{$\sigma_{\rm v}$}	& \multicolumn{1}{c}{$P$}  & \multicolumn{1}{c}{TiO}  & \multicolumn{1}{c}{ZrO}\\
   \ & \multicolumn{1}{c}{(J2000.0)} & \multicolumn{1}{c}{(J2000.0)} & \multicolumn{1}{c}{Type}		& \multicolumn{1}{c}{$K_s$}	& \multicolumn{1}{c}{($J-K_{\rm s}$)}	& \multicolumn{1}{c}{(km s$^{-1}$)}	& \multicolumn{1}{c}{(km s$^{-1}$)}	& \multicolumn{1}{c}{\%}	& \multicolumn{1}{c}{index}	  & \multicolumn{1}{c}{index}\\
 \hline
0001 &  18 55 12.44 &    --30 23 39.0  &   K   &     9.92 &  0.35  &   29.53 & 12.19 & 0.00 & 1.04 & 0.91\\
0002 &  18 55 18.97 &    --30 21 01.7  &   K   &     9.26 &  0.75  & --14.93 &  1.24 & 0.00 & 1.03 & 0.95\\
0003 &  18 54 12.34 &    --30 24 10.0  &   K   &    12.28 &  0.78  &  --8.96 &  1.68 & 0.00 & 1.01 & 0.96\\
\nodata & \nodata & \nodata & \nodata & \nodata & \nodata & \nodata & \nodata & \nodata & \nodata & \nodata \\
\hline
\end{tabular}
\end{minipage}
\end{table*}

\begin{figure*}
 \resizebox{\hsize}{!}{\includegraphics[angle=270]{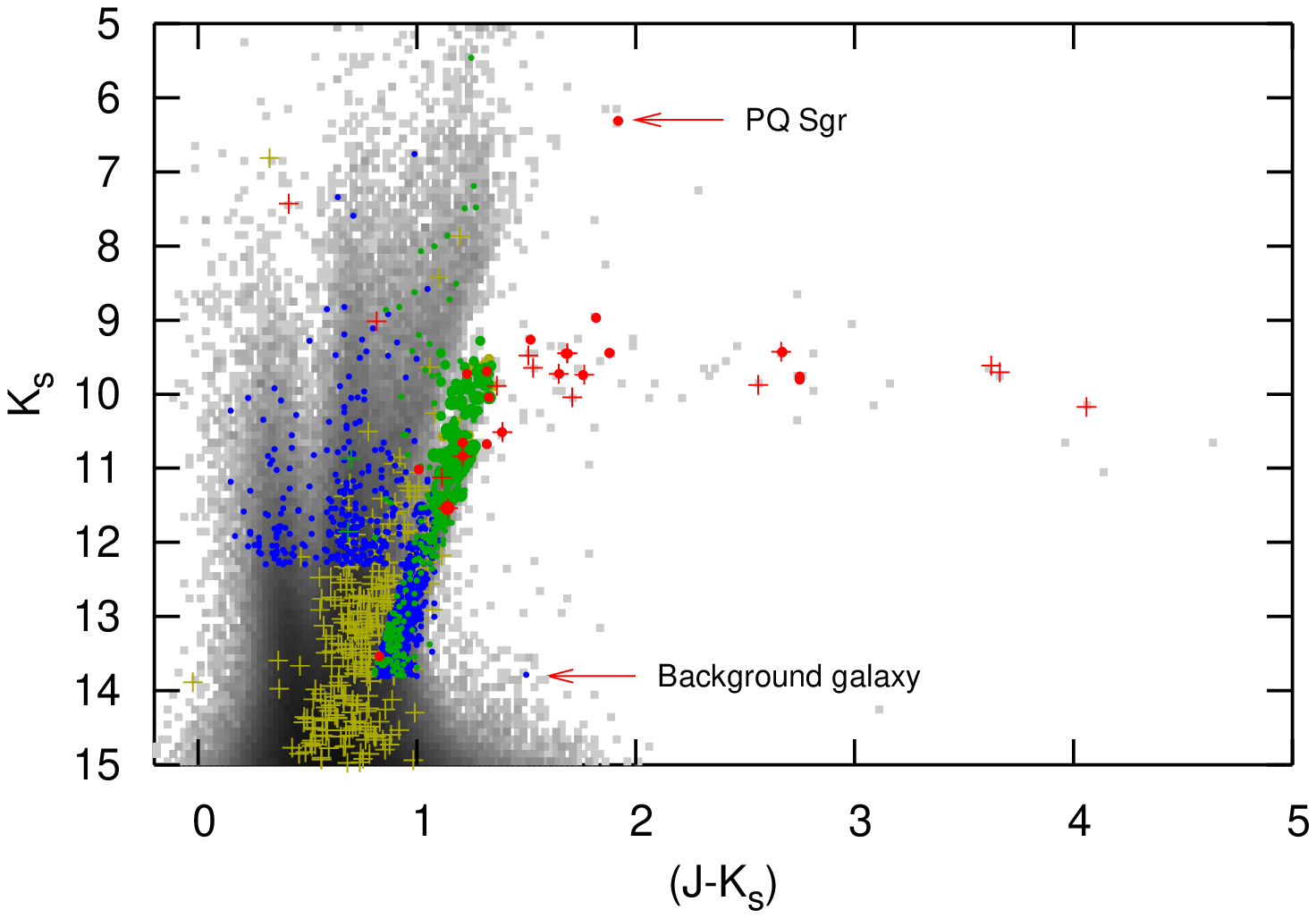} \qquad \includegraphics[angle=270]{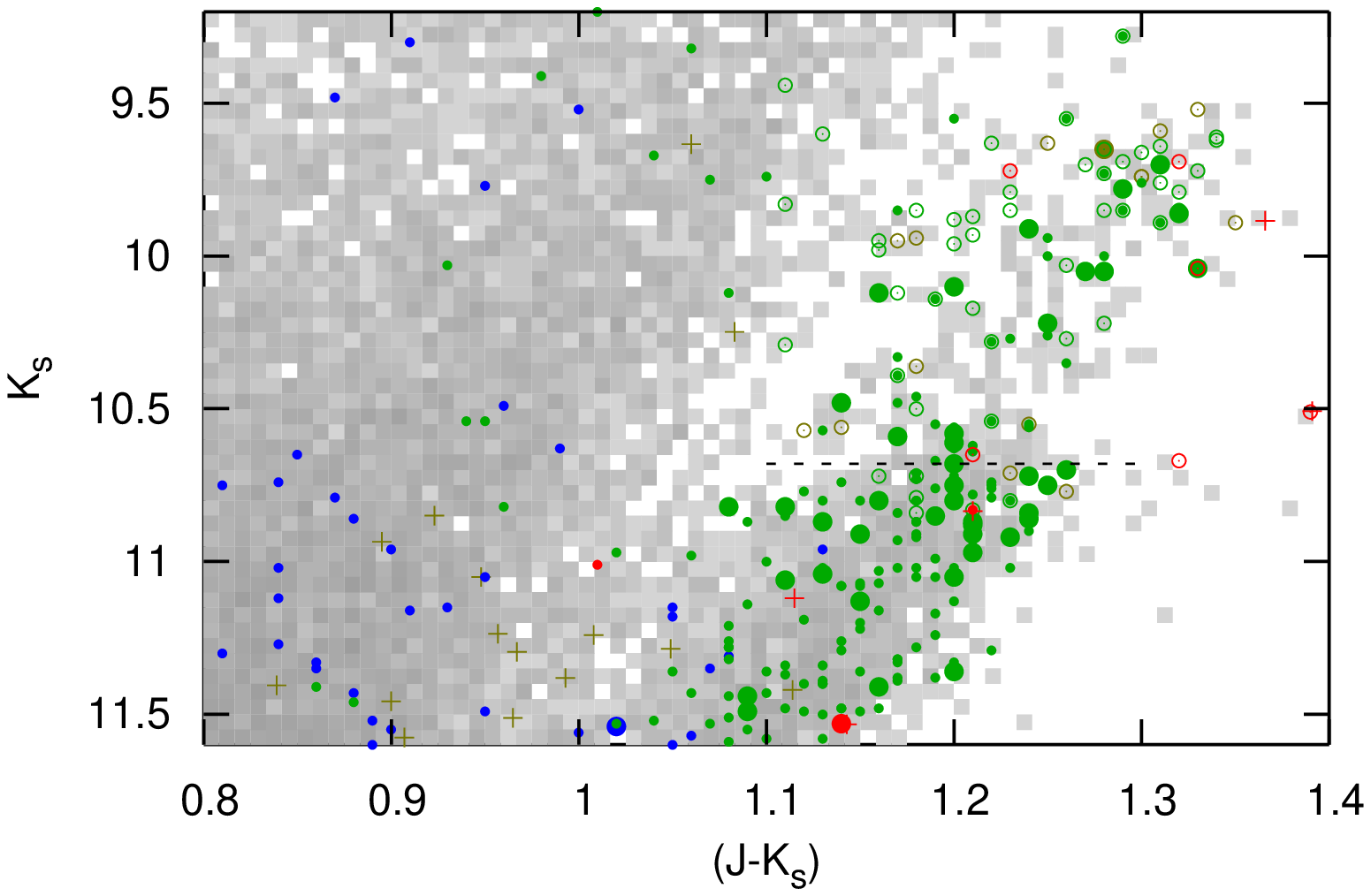}}
 \resizebox{\hsize}{!}{\includegraphics[angle=270]{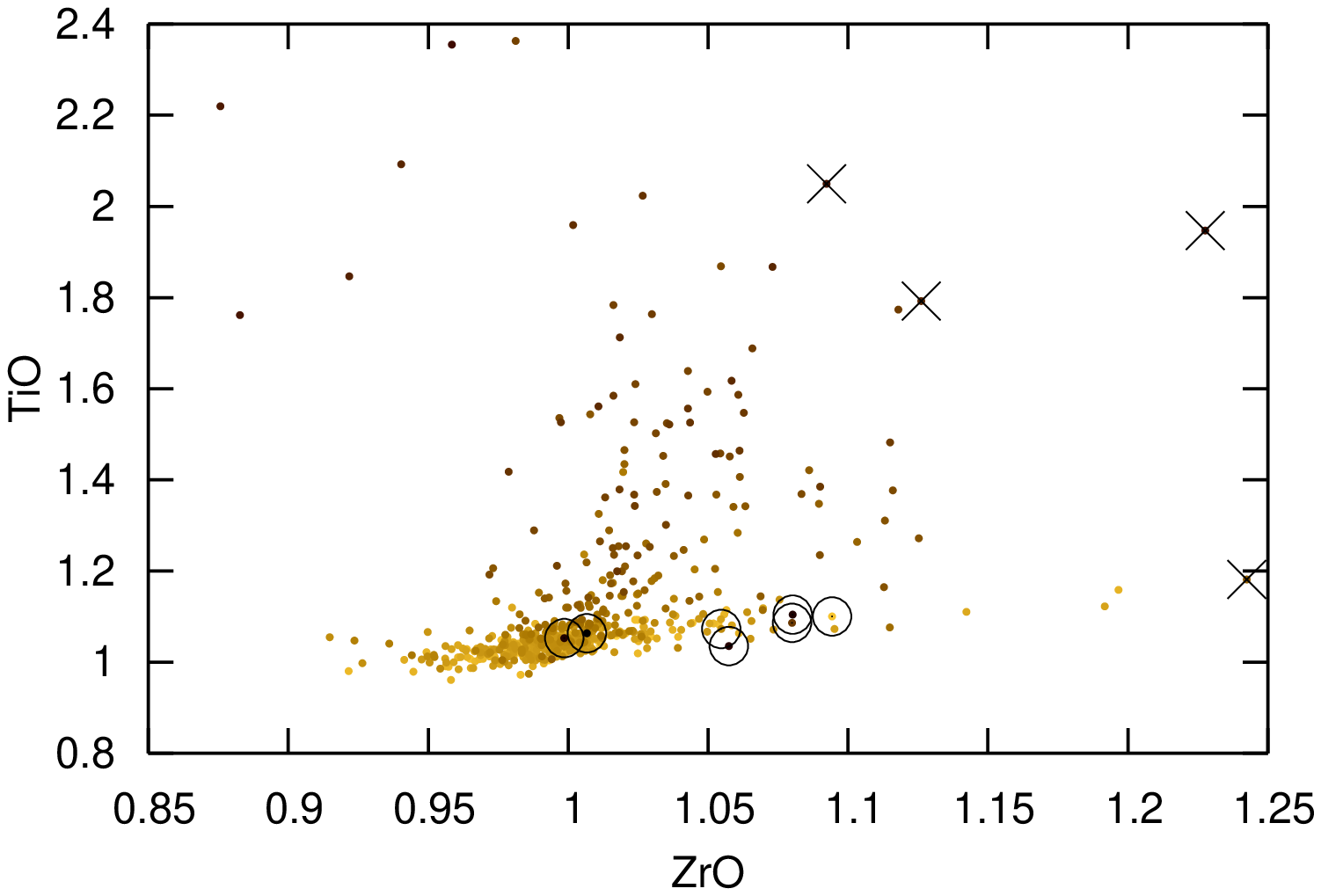} \qquad \includegraphics[angle=270]{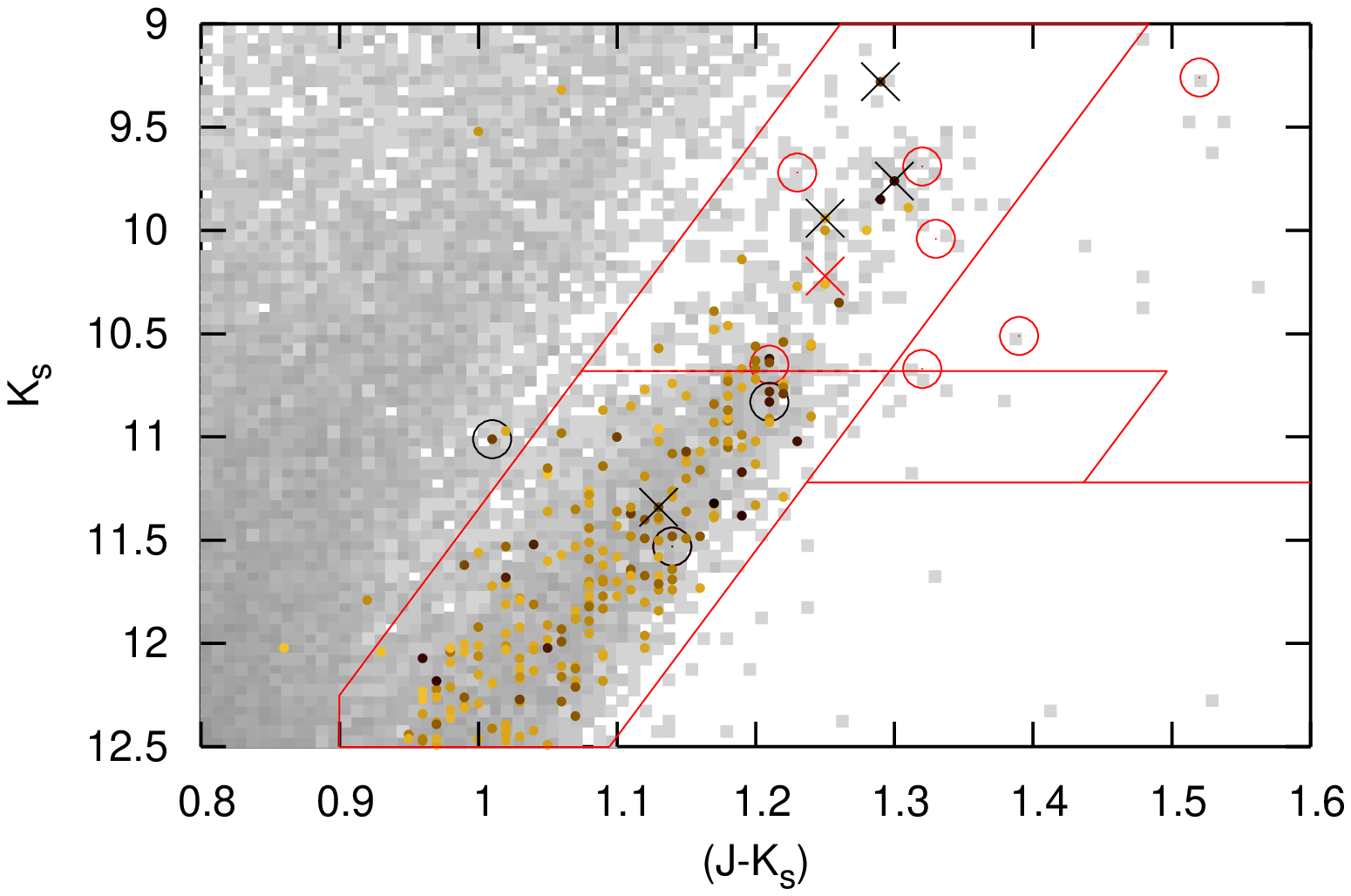}}
 \caption{\emph{Top panels:} Colour--magnitude diagrams showing the distribution of spectral types in the observed sample (circles) and literature (plus signs). Colours represent: (red) carbon stars, (green) M giants, (blue) earlier spectral types, (yellow) undetermined or unlisted spectral type. Hollow symbols show SAAO targets, smaller filled symbols show FLAMES targets, larger symbols show UVES targets. The observed background galaxy and Galactic carbon star PQ Sgr are labelled. The right-hand panel shows a zoomed-in region of the left-hand panel near the RGB tip, which is marked by the horizontal line.\newline
\emph{Bottom panels:} Molecular band strengths of the GIRAFFE sample, using indices from \citet{OGG11}. Carbon stars are shown by circles, the S stars by crosses. \emph{Bottom-left panel:} The colour scale shows ($J-K_{\rm s}$) colour, with darker stars being redder. \emph{Bottom-right panel:} Colour--magnitude diagram, showing the colour cuts applied from Figure \protect\ref{CutsFig}. The colour scale shows the strength of the ZrO band index. Carbon and S-type stars in the UVES and SAAO samples are shown as lighter (red) points.}
 \label{TypeFig}
\end{figure*}

An initial visual inspection of each spectrum was performed to separate carbon- and oxygen-rich stars for the purpose of radial velocity determination. This was done on the basis of C$_2$/CN and TiO/VO bands, respectively (see Figure \ref{SpecFig}). The final spectral types are listed in Tables \ref{VLTTargets} and \ref{SAAOTargets}. They are displayed along with literature results in Figure \ref{TypeFig}.

VLT spectra without discernable molecular bands were assumed to be of spectral type K. SAAO spectra with no discernable bands were labelled as unknown spectral type (U). The difference in approach arises from the different data qualities. Due to poor observing conditions during the 2010 observing run, some SAAO data are of such low signal-to-noise in some cases that even significant molecular bands would not be visible. The VLT spectra are of sufficient signal-to-noise that TiO bands would always be visible if they are present. The C$_2$ and CN features of carbon stars are not as clear as TiO features in the VLT data, meaning some carbon stars could theoretically be misclassified as K stars. However, the stars we classify as K stars are all at low luminosities where only a few extrinsic carbon stars are expected, so we believe we have listed all observed intrinsic carbon stars in our sample.

One galaxy (\#29) is observed in the sample. The spectrum appears featureless, but optical and 2MASS images of the source show that it is elongated. We therefore discount it from further analysis. This leaves us with 718 K-type, 290 M-type and 7 C-type stars observed with GIRAFFE, respectively; 1, 41 and 1 with UVES; and $\leq$15, 54 and 15 with the SAAO 1.9-m. The higher fraction of carbon stars in the SAAO data is an intentional bias, introduced by preferentially observing the brightest AGB stars.



\subsection{Radial velocity determination}

Radial velocities for the VLT/GIRAFFE targets were determined using the Starlink routine {\sc hcross} and corrected from topocentric to barycentric co-ordinates using the task {\sc rv}. We used the stars 1013, 1014 and 1015 were used as templates for the K, M and C stars, respectively (Figure \ref{SpecFig}). The three templates were visually cross-correlated with a 4000 K {\sc marcs} model with $\log g = 1.5$ dex, [Z/H] = --0.75 and [$\alpha$/Fe] = +0.3 \citep{GBEN75,GEE+08,MvLD+09} to provide absolute radial velocities with an estimated error of 2 km s$^{-1}$.

Each star was cross-correlated against its relevant template over 30 spectral regions across the four spectral settings. These spectral regions were chosen to bracket particular features whilst avoiding areas of telluric absorption. Values lying more than 2.17$\sigma$ (97\% confidence interval) from the median were iteratively clipped, and a single radial velocity for each star was determined using the median of these remainder. The resulting random error in radial velocity averages 1.8 km s$^{-1}$, with 87\% of sources having errors $<$10 km s$^{-1}$. The remaining 13\% of sources are mostly of too low signal-to-noise to accurately determine a radial velocity.

Lacking an absolute radial velocity standard, we assume that the average Sgr dSph star (those with $100 < v_{\rm r} < 200$ km s$^{-1}$ and $\sigma_{\rm v} < 10$ km s$^{-1}$) has a radial velocity of $v_{\rm r} = 141$ km s$^{-1}$ (cf.\ \citealt{IWG+97,ZW96,Harris96,GSZ+10,PZI+11}), and offset our velocities accordingly. Only one of our stars has a published radial velocity (2MASS 18554672--3035248, our source \#346; \citealt{CBG+10}, their source 23001277, $v_{\rm r} = 147.6$ km s$^{-1}$). Our velocity (148.22 $\pm$ 1.05 km s$^{-1}$) is in agreement within the errors, showing that our absolute velocity calibration is accurate.

The radial velocities of the 43 observed UVES targets were determined using a similar method. Here, however, radial velocities could not be calculated using the Starlink {\sc hcross} package. This appears to be due to problems caused to the Fourier transformed data by poor order matching, increased noise in the spectrum, and the generally richer and more-varied spectra of these typically more-evolved stars. To combat this, we removed small-scale structure in the spectra by smoothing them by a Gaussian of $\sigma = 10$ pixels and removed large-scale structure by dividing them by the same spectrum smoothed by a Gaussian of $\sigma = 100$ pixels). Each spectrum was sliced into seven segments, each of 3000 pixels, and each segment was cross-correlated with the corresponding segment from every other star. As above, an iteratively-clipped (2$\sigma$) median was used to calculate a unique radial velocity and the standard deviation of the measurements used to create a measurement of the velocity error.

Of the resulting 43 radial velocities, 42 are very tightly grouped and obviously belong to the Sgr dSph. The 43rd star, \#1033, is Galactic. We adjust the radial velocity zero point so that the average of the 42 Sgr dSph stars lies at 141 km s$^{-1}$ to place them on a common scale with the GIRAFFE spectra.

Accurate radial velocities could not be determined for the SAAO data due to the low spectral resolution and poor data quality. 

\subsection{Membership probability}
\label{MemSect}

To define a membership probability of the Sgr dSph, we approximate the 556 stars with Sgr-dSph-like velocities (as defined above) and 312 Galactic foreground stars (those with $-150 < v_{\rm r} < 100$ km s$^{-1}$ and $\sigma_{\rm v} < 10$ km s$^{-1}$) as two Gaussians, with $<\!v_{\rm Sgr}\!> = 141.0$ and $\sigma_{\rm Sgr} = 12.7$ km s$^{-1}$, and $<\!v_{\rm Gal}\!> = -10.3$ and $\sigma_{\rm Gal} = 45.3$ km s$^{-1}$, respectively. Membership probability of the Sgr dSph ($P$) was then defined as:
\begin{equation}
P = \frac{\displaystyle\int_v \frac{556 \exp\left(-\frac{(v-<v_{\rm Sgr}>)^2}{\sigma_{\rm Sgr}^2}\right)}{312 \exp\left(-\frac{(v-<v_{\rm Gal}>)^2}{\sigma_{\rm Gal}^2}\right)} \cdot \exp\left(-\frac{(v-v_{\rm r})^2}{\sigma_{\rm v}^2}\right) \mathrm{d}v}{\frac{312}{556}\int_v \exp\left(-\frac{(v-v_{\rm r})^2}{\sigma_{\rm v}^2}\right) \mathrm{d}v} ,
\end{equation}
where $v$ is a test radial velocity, $v_{\rm r}$ is the calculated (clipped median) radial velocity and $\sigma_{\rm v}$ is its error.

\begin{table}
  \caption{Spectral types of targets observed with the SAAO Radcliffe 1.9-m. A full table is available in the online version.}
\label{SAAOTargets}
  \begin{tabular}{lc@{\ \ }c@{}c@{}cc}
  \hline\hline
   \multicolumn{1}{c}{ID}	& \multicolumn{1}{c}{RA}	& \multicolumn{1}{c}{Dec}	& \multicolumn{1}{c}{Spectral}	& \multicolumn{2}{c}{2MASS} \\
   \ & \ & \ & \multicolumn{1}{c}{Type}		& \multicolumn{1}{c}{$K_s$}	& \multicolumn{1}{c}{($J-K_{\rm s}$)} \\
  \hline
2001  &  18 56 11.24  &   --29 03 54.9  &   C   &    06.31 &  1.92 \\
2002  &  18 53 59.53  &   --29 58 57.4  &   C   &    09.26 &  1.52 \\
2003  &  18 54 09.02  &   --30 23 20.2  &   M   &    09.28 &  1.29 \\
\nodata & \nodata & \nodata & \nodata & \nodata & \nodata \\
  \hline
\end{tabular}
\end{table}

\begin{figure}
 \resizebox{\hsize}{!}{\includegraphics[angle=270]{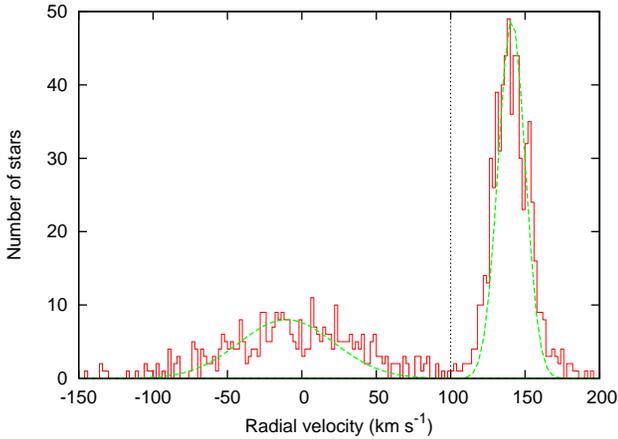}}
 \caption{Histogram of derived radial velocities of FLAMES targets. The dashed curves are Gaussians fit to the Sgr dSph and Galactic populations, as described in \S\ref{MemSect}. The Galactic population is relatively well separated, lying almost exclusively below 100 km s$^{-1}$ (vertical line).}
 \label{RVHistFig}
\end{figure}

\begin{figure}
 \resizebox{\hsize}{!}{\includegraphics[angle=270]{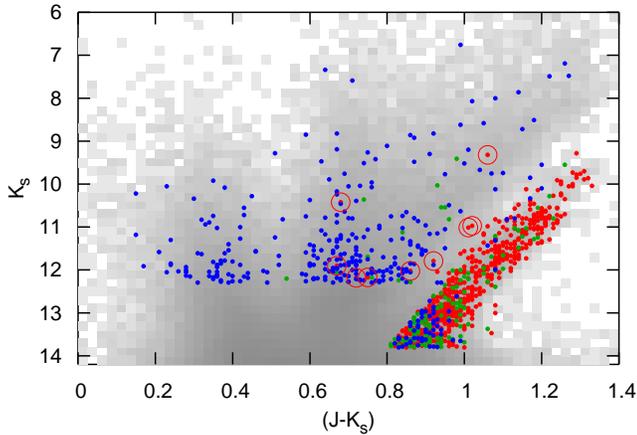}}
 \caption{Colour--magnitude diagram showing the positions of Sgr dSph members (red), Galactic foreground stars (blue) and stars of indeterminate location (green) observed with the VLT, overlaid on a density plot of 2MASS sources on a logarithmic greyscale. Circled objects highlight stars with velocities co-incident with the Sgr dSph, but colours that are not. These may be metal-poor or post-AGB member stars.}
 \label{CMDFig}
\end{figure}

Figure \ref{RVHistFig} shows a histogram of these velocities. A discernable break is apparent between Galactic stars and Sgr dSph members at $v_{\rm helio} \approx +100$ km s$^{-1}$, beyond which most stars are Sgr dSph members. The velocity selected (non-)members and stars whose membership could not be reliably determined are shown in Figure \ref{CMDFig}, which shows that the Sgr dSph members clearly lie almost exclusively along the redder giant branch. Based on our probability analysis, we find 593 stars are members of the Sgr dSph at high ($>$95\%) confidence, 338 stars are members of our Galaxy, and 127 stars could not reliably be placed in either category.

\subsection{Band indices and S stars}
\label{BandSect}


Having a spectrum in the stellar rest frame, we can now perform more accurate spectral analysis. To do this, we adopt the TiO and ZrO indices from \citet{OGG11}, inverting the TiO index as per their figure 2. Namely, these are the ratio of the average flux at 640--646 nm to that at 647.5--653.5 nm for the ZrO band, and 696.5--702.8 nm to 706.5--717.5 nm for the TiO band. We do not cover the spectral region required to calculate the CT51 index of \citet{OGG11}, so we cannot repeat their analysis exactly. 

Figure \ref{TypeFig} shows the band indices we derive. Objects with $I_{\rm ZrO} > 1.05$ were visually searched for ZrO bands. We identified S stars by comparison to carbon-enriched spectra from \citet{vENP+11}. As these spectra are not yet fully published, a visual comparison to their Figure 2 was necessary. Four M stars were reclassified as S stars identified on this basis: \#361, 424, 909 and 942. A fifth star from the UVES spectra (\#1054) was also reclassified as an S star. We return to these stars in Section \ref{SStarSection}.

The remainder of the stars cluster close to unity, with the cooler, M-type, stars displaying increasingly large values of the TiO index. The carbon stars, although equally cool, do not share the TiO or ZrO features, thus are the only very cool stars to appear near unity in both indices.




\section{Discussion}

\subsection{Velocity gradients}

\begin{figure}
\resizebox{\hsize}{!}{\includegraphics[angle=270]{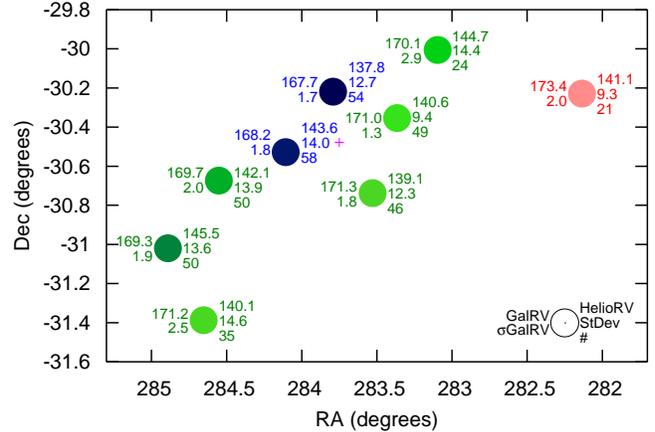}}
 \caption{Average radial velocity of K stars in the observed fields. Labels (as shown in the bottom right) indicate (left) the average radial velocity in the Galactic standard of rest (also coloured dark blue to light red) and its error, and (right) the average heliocentric velocity, its standard deviation and number of members observed. The small magenta plus sign marks the location of M54.}
 \label{VgradFig}
\end{figure}

To investigate radial velocity gradients in the Sgr dSph, we have taken the K star population in each observed field and computed an average velocity for those K stars with velocity uncertainties under 10 km s$^{-1}$ and corrected these back to the Galactic standard of rest using the NASA/IPAC Extragalactic Database (NED) velocity converter\footnote{http://ned.ipac.caltech.edu/help/velc\_help.html}. We show these in Figure \ref{VgradFig}. A possible north-east--south-west gradient of $\pm\sim$2 km s$^{-1}$ exists, which could conceivably be aligned with the galaxy's minor axis and/or parallel to the Galactic plane, however it is at the limit of detection and therefore cannot be conclusively shown. Neither can we determine on the basis of these data alone whether such a gradient is indicative of a tidal distortion or a true rotation.

\subsection{Investigating the metal-poor population}

Figure \ref{CMDFig} shows nine objects with a high membership probability ($>$99\%), but with ($J-K_{\rm s}$) colours much bluer than the main Sgr dSph population, namely \#10, \#66, \#105, \#155, \#170, \#190, \#232, \#243 and \#260. Given the slight overlap between the Galactic and Sgr dSph velocity distributions (Figure \ref{RVHistFig}), we cannot be certain that these are Sgr dSph stars based on their velocities alone. These stars are important, as they may be members of the metal-poor population, or post-AGB stars.

Following a thorough visual investigation of their spectra and associated literatue data, we are unable to conclusively determine (non-)memberships of these stars. However, we suggest that \#155 and \#260 are probably Sgr dSph stars, based on their proximity to the Sgr dSph branch on the ($J$--$K_{\rm s}$) colour--magnitude diagram, and on the carbon-richness of \#260. We further suggest that \#66, \#105 and probably \#243 are Galactic, based on their proper motions \citep{RDS10}. The other stars remain at an undetermined distance, as we have insufficient data to determine their location without a full spectral synthesis to determine their chemical abundances. Removing \#66, \#105 and \#243 from the Sgr dSph sample has little effect on our overall results (entirely within the measurement errors). The most notable change is that removing \#243 from the most-northerly field decreases its average heliocentric radial velocity by 0.9 km s$^{-1}$ and its standard deviation by 0.4 km s$^{-1}$ (cf.\ Figure \ref{VgradFig}).



\subsection{The stellar evolution rate}
\label{EvolRateSect}

\subsubsection{Determining evolutionary rates}
\label{DEvolRateSect}

\begin{table*}
  \caption{Star counts from the regions shown in Figure \protect\ref{CutsFig}.}
\label{CountsTable}
  \begin{tabular}{@{}l@{}r@{}r@{}rr@{}rrrrr@{}r@{\ \ }r@{\ \ }r@{}}
  \hline\hline
   \multicolumn{1}{c}{Regi\rlap{on}}	&    \multicolumn{2}{c}{Total count}	&    \multicolumn{2}{c}{Confirmed}	&    \multicolumn{1}{c}{Carbon star}	&    \multicolumn{3}{c}{r.v.\ member?}	&    \multicolumn{4}{c}{Implied number in core}\\
   \multicolumn{1}{c}{\ }	&    \multicolumn{1}{c}{Body$^1$}	&    \multicolumn{1}{c}{Core$^1$}	&    \multicolumn{1}{c}{M-type}	&    \multicolumn{1}{c}{Carbon}	&    \multicolumn{1}{c}{fraction$^2$}	&    \multicolumn{1}{c}{Yes}	&    \multicolumn{1}{c}{No}	&    \multicolumn{1}{c}{Fraction$^2$}	&    \multicolumn{2}{c}{Carbon stars$^3$}	&    \multicolumn{2}{c}{Members$^3$}\\
  \hline
(a)   & 7327 & 2038 & 182 & 2 & 1.3 $\pm$ 0.7\%  & 234 & 7 & 97.0 $\pm$ 0.9\% & 12--37 & (12--38) & 1959--1995 & (1915--2039)\\
(a)$^4$&2385 &  700 & 119 & 2 & 2.0 $\pm$ 1.0\%  & 114 & 4 & 96.4 $\pm$ 1.4\% &  6--20 &  (5--20) &   665--685 &   (640--710)\\
(b)   &  563 &  162 &  44 & 4 & 9.1 $\pm$ 3.4\%  &  38 & 3 & 92.1 $\pm$ 3.5\% &  4--14 &  (4--15) &   144--155 &   (132--167)\\
(c)   &  161 &   35 &   0 &14 & 98.0 $\pm$ 2.6\% &   4 & 0 & 93.9 $\pm$ 8.2\% & 35$^5$ & (29--41) &     35$^5$ &     (29--41)\\
(e)   &   17 &    8 &   0 & 0 &     \nodata      &   0 & 0 &     \nodata      &   0--8 &  (0--11) &       0--8 &      (0--11)\\
(f)   &  993 &  301 &  57 & 1 & 2.4 $\pm$ 1.6\%  &  55 & 2 & 96.0 $\pm$ 2.1\% &  2--11 &  (2--12) &   283--295 &   (267--312)\\
(g)   &  216 &   56 &  16 & 0 & 2.5 $\pm$ 2.3\%  &  13 & 1 & 91.2 $\pm$ 6.1\% &   0--3 &   (0--3) &     48--54 &     (41--62)\\
  \hline
\multicolumn{13}{p{0.95\textwidth}}{$^1$Defined in Figure \protect{\ref{AGBFig}}. $^2$The error is defined using the probability and error approximations of \protect\citet{Wilson27}. $^3$The left column of each pair denotes the implied number, the right column (bracketted) includes the Poisson noise necessary to accurately determine evolutionary rates. $^4$Only stars above $K_{\rm s} = 11.533$ mag. $^5$All stars in this region are assumed to be carbon stars within the Sgr dSph.}
\end{tabular}
\end{table*}

One can, to first order, estimate the rate at which stars evolve off the AGB by comparing the observed colour--magnitude diagram to stellar evolutionary tracks. RGB evolutionary models tend to be more easily-available and much better calibrated than those of the AGB. Furthermore, many evolutionary models (including the models used here) do not include mass loss, or assume a prescription of mass loss which may not be appropriate. For stars above $\sim$1 M$_\odot$, RGB mass loss is relatively small and can usually be ignored, and (as we will show) RGB evolutionary rates have little dependence on stellar mass in most normal regimes. It is therefore easier to use RGB models to determine the RGB evolution rate, then determine the AGB evolution rate by using the ratio of star counts on the RGB and AGB. This should be identical provided every RGB star eventually ends its AGB evolution above the RGB tip.

Throughout this process, we will refer to the regions labelled (a) through (g) in Figure \ref{CutsFig}. Star counts for these regions are displayed in Table \ref{CountsTable}. In this Figure, region (a) corresponds to the unreddened RGB, region (b) to the unreddened AGB and region (c) to the reddened AGB. Region (d) is largely devoid of interesting stars, containing mostly background galaxies. Region (e) may correspond to a reddened RGB, though the true identity of these stars (be they RGB or AGB, carbon- or oxygen-rich) is unknown. Regions (f) and (g) show the RGB and AGB populations we use to determine evolutionary rates. Note that these regions, each covering $\Delta K_{\rm s} = 0.4$ mag, avoid the 0.1 $K_{\rm s}$-band magnitudes either side of the RGB tip (at $K_{\rm s} = 10.68$ mag) to account for uncertainties in the 2MASS photometry and uncertainty in the position of the RGB tip itself.

Table \ref{CountsTable} also contains four columns which extrapolate the number of carbon stars and Sgr dSph members to the entire $3^\circ \times 3^\circ$ core region (as defined in Figure \ref{AGBFig}). In each case, we show two numbers. The first is the 68\% confidence interval for the number in that region, based on multiplying the relevant observed fraction (carbon star fraction or membership fraction) by the raw counts. The second shows the same confidence interval taking into account the Poissonian error in the number of stars. This latter correction is necessary when exploring evolutionary rates, as we see an instantaneous measure of the number of stars in a particular evolutionary phase ($n$), which should vary over time by $\sim\sqrt{n}$.

The values in Table \ref{CountsTable} are subject to biases in our detection statistics. These biases are dominated by three key factors:
\begin{enumerate}
\item The assumption of homogeneity across the core of the Sgr dSph, and that the observed GIRAFFE fields are representative of the core region. We see no obvious difference among the stars observed in the nine different VLT fields, either in the membership fraction of the individual regions in Table \ref{CountsTable}, nor the fraction of carbon stars in each field\footnote{We note that the overall fractions may vary from field to field due to the differing contamination by foreground stars, but this has little effect on our colour-selected samples.}. We note that M54, which comprises of a population of different metallicity, lies within the core region, but this is represents a small ($\approx$4--5\%) contribution to the total number of objects. We have not sampled M54 spectroscopically, so it should only have a secondary effect on the fractions we list in Table \ref{CountsTable}, but it does make a contribution to the total number of objects.
\item The sampling of the GIRAFFE spectra within each field. While observations were limited by the placement of fibres, we have no reason to believe that the precise position within a field should bias our spectra. FLAMES targets were essentially chosen randomly from sources within regions (a)--(e) within the constraints of the GIRAFFE field. The UVES sample is slightly biased towards brighter targets, but these only represent a few percent of our targets and, because we can be confident of the spectral types and radial velocities of these stars, should not lend a significant bias to the results presented in Table \ref{CountsTable}.
\item The difference in observation characteristics and spectral typing efficiency in the SAAO observations. As no radial velocities are measured for these stars, this can only bias the carbon star fraction. Within this, there should only be a significant impact on the fraction in region (b), as we assume all stars within region (c) are carbon stars and there are only eight stars successfullly observed by SAAO in region (a). While our detection efficiency of carbon stars in region (b) may be biased due to the variable signal-to-noise of these observations, the small number of carbon stars this region can realistically host means any bias can have little impact on the total number of carbon stars in the core region. Removing the SAAO spectra from the analysis leads to an increased error budget (primarily due to the absence of carbon stars in region (b)), but no substantial changes to the timescales involved, as the error budget is already quite large.
\end{enumerate}
We therefore do not expect biases in our data to affect the timescales we find by greater than a few percent and have included a conservative 5\% error in our timescales to account for this.

\subsubsection{The stellar death rate}
\label{DeathRateSect}

The first step in our calculation is to determine the evolutionary rate near the top of the RGB in kyr mag$^{-1}$. The Dartmouth evolutionary tracks \citep{DCJ+08} contain a pre-computed model grid spaced at $\Delta$[Fe/H] = 0.5 dex and $\Delta M$ = 0.05 M$_\odot$. We assume solar [He/H]. Interpolating over this grid, we identify the time taken for a star to traverse the upper-RGB region (f). Using bounds of --0.7 $<$ [Fe/H] $<$ --0.4 dex, 4 $<$ age $<$ 8 Gyr and assuming [$\alpha$/Fe] = +0.2 dex, we find that stars should traverse region (f) over a period between 1020 and 1106 kyr.

The next step is to determine the number of Sgr dSph stars that exist in this region. We focus on the galaxy's core (where our observations were taken) to determine numbers as our membership selection efficiency will vary across the galaxy (see Figure \ref{AGBFig} for the definition of areas). Table \ref{CountsTable} shows we estimate that 267--312 Sgr dSph stars in region (f) of the galaxy's RGB.

We must now remove the contaminating AGB from the RGB sample. We can estimate the number of these by counting the stars in the region above the RGB tip (region (g)), which should all be AGB stars. Table \ref{CountsTable} lists 41--62 Sgr dSph AGB members in this region. Typically, models (including the aforementioned Dartmouth models) predict that the rate of evolution up the AGB increases with luminosity, however the observed $K_{\rm s}$-band luminosity function between $9.58 < K_{\rm s} < 10.58$ mag is flat (--11\% $\pm$ 20\% increase in number per magnitude fainter in $K_{\rm s}$). We can therefore expect a similar number of AGB stars to pollute the RGB sample, albeit with an increased uncertainty due to the slope of the AGB luminosity function. We calculate that 32--65 Sgr dSph AGB stars pollute the RGB sample, and therefore expect that region (f) contains 202--280 Sgr dSph RGB stars.

Based on the Dartmouth models' evolutionary rate through this period, we expect one star to evolve off the RGB every 3500--5700 years in the galaxy's core. We expect the majority of these stars to subsequently evolve through their horizontal branch and AGB evolution. Given it takes a zero-age horizontal branch star only $\sim$100--200 Myr to complete its AGB evolution as far as the RGB tip \citep{MGB+08}, we can say that the rate of AGB evolution (in units of years per star) is similar to the RGB. Therefore, one star in the core $3^\circ \times 3^\circ$ region also evolves off the AGB every 3500--5700 years. Assuming evolutionary rates in the rest of the galaxy are similar (which is only true to first order due to the radially-changing age and metallicity of the population) then we can simply scale by the number of sources in regions (a--c) to find the evolutionary rate for the entire main body of the galaxy. Under these assumptions, a star would evolve off the AGB every 1000--1700 years somewhere in the main body of the galaxy (we remind the reader that the main body is defined in this context in Figure \ref{AGBFig}).

\subsubsection{The thermal-pulsing and optically-obscured lifetimes}
\label{TPTimeSect}

We can further note that the upper AGB (above the RGB tip; regions (b) and (c)) contains $\approx$183 \emph{bona fide} Sgr dSph AGB stars (core region only). Using the above timescale per star, we can estimate the time a star spends on the upper AGB to be between 630 and 1060 kyr, on average. Due to the spread in stars' initial properties, this is likely to vary quite significantly, probably by more than the error budget. The total extent of the upper AGB is $\Delta K_s \approx 1.2$ mag. If we assume that AGB evolution is 3.26--6.83$\times$ (viz.\ 280--202 RGB stars / 41--62 AGB stars) faster than RGB evolution which, according to the Dartmouth models, takes 1020--1106 kyr per 0.4 mag, this sets a duration to the upper-AGB lifetime of 430--1070 kyr. Similarly, if we assume that thermal pulses begin $\approx$0.3 $K_{\rm s}$-band magnitudes below the RGB tip (in accordance with the above models) then the same evolutionary rate (in terms of kyr mag$^{-1}$) gives a TP-AGB lifetime is 530--1330 kyr.

Typical thermal pulse intervals for solar-mass AGB stars are $\sim$100 kyr: \citet{KL07} determine 143 kyr for a 1 M$_\odot$, [Z/H] = --0.68 model including mass loss, and 75 kyr for a 1 M$_\odot$, [Fe/H] = --0.38 which does not include mass loss. Taking a range of 75--150 kyr implies that these stars undergo 3--14 thermal pulses on the upper AGB, or 4--17 thermal pulses on the TP-AGB in total. For comparison, \citet{MG07} predict a total of 11 thermal pulses for a 1.0 M$_\odot$, [Z/H] = --0.38 model.

Similarly, we can estimate the time a star spends in the optically-obscued carbon-rich phase by simply counting the stars in region (c). This region contains 35 stars, of which we believe all are carbon-rich members and all of which are sufficiently reddened in $(J-K_{\rm s})$ to be denoted as optically-obscured. Accounting for the Poisson error in this count, we find the average lifetime of this phase to be 100--235 kyr.

These figures should be treated as approximations, however, for two reasons:
\begin{enumerate}
\item We assume that net circumstellar absorption by dust is negligible in the $K_s$ band. The reddened, carbon-rich AGB extends slightly further above the AGB in $K_s$-band, and may extend somewhat further in bolometric luminosity, but contains relatively few stars, meaning stars do not typically survive long in this phase. This could potentially increase the maximum upper-AGB lifetime by another $\sim$25\%, based on relative star counts between regions (b) and (c). This would give a maximum of $\lesssim$17 thermal pulses on the upper AGB, or $\lesssim$21 thermal pulses on the total TP-AGB.
\item We assume all stars within our colour cuts belong to the bulk population. Some will be part of the (very) metal-rich populations, though the number of such stars currently on the RGB and AGB is a factor of several smaller than the number in the bulk population. The metal-poor populations are largely unaccounted for here: the giant branch is well-hidden within the foreground Galactic population. Only significantly-reddened stars (of which there will be very few) from this population will enter our colour cuts. This could change the ratio of stars in the different regions, hence our relative timescales, but only by a factor much less than the sub-populations contribution (by fraction) to the total population. We therefore estimate this will have at most a few percent effect on our results.
\end{enumerate}
As our error budget is quite large, the net effects of these assumptions have little effect on our overall timescales.

\subsection{Which stars become carbon-rich?}
\label{COSection}

\begin{figure}
\resizebox{\hsize}{!}{\includegraphics[angle=270]{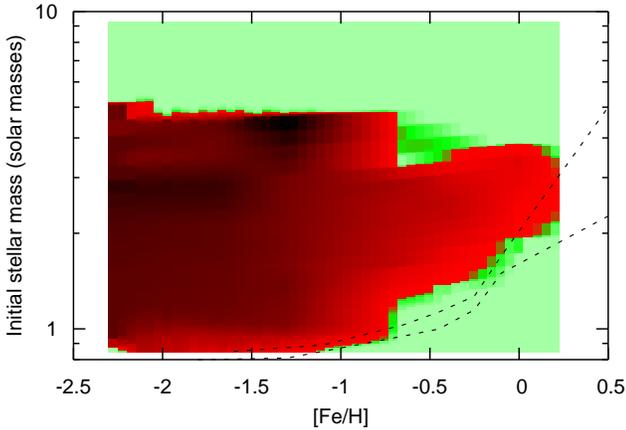}}
 \caption{Maximum C/O ratio attained by stars at a given metallicity and initial mass, based on isochrones from \citet{MG07}. Light green areas show regions where C/O never exceeds unity; red areas show regions where stars become carbon rich, with darker colours representing increasing values of the maximum C/O ratio. The dashed lines show the ranges found in the Sgr dSph, based on \protect\citet{SDM+07} and using \protect\citet{MG07} to covert ages into masses.}
 \label{MaxCOFig}
\end{figure}


The origin of the Sgr dSph's carbon-star population has so far been unclear. Which of the populations do carbon stars come from? We have identified 23 carbon stars in our spectra (Tables \ref{VLTTargets} \& \ref{SAAOTargets}), all but four of which are new\footnote{\#705 = WMIF 4 \protect{\citep{WMIF99}}; \#1044 = WMIF 10; \#1015 = WMIF 12; \#750 = Sgr 09 \protect{\citep{LZS+09}}.}. This more than doubles the number of confirmed carbon stars in the galaxy. The brightest identified carbon star, \#2001, is the mass-losing star PQ Sgr ($P$=204 d, ASAS; \citealt{Pojmanski97}). As it was observed with SAAO, it has no radial velocity measurement. Its location on the colour--magnitude diagram, however, suggests that it is probably a Galactic star: it is the brightest carbon star in Figure \ref{TypeFig} and sits three $K_s$ magnitudes above the carbon-rich AGB.

Figure \ref{MaxCOFig} shows the region in mass--metallicity space where stars are expected to become carbon-rich, according to the theoretical isochrones of \citet{MG07}. This region is bounded at high mass by hot bottom burning, which destroys carbon which would otherwise enhance the surface C/O ratio \citep{SDU75,DAM96,VM10}. More importantly, it is bounded at low mass and high metallicity by the amount of dredge-up that occurs: low mass stars have little dredge-up, which can prevent C/O reaching unity; high metallicity stars have high natal oxygen abundance, meaning more carbon needs dredged up to increase C/O to unity.

All three populations of the Sgr dSph are near this low-mass--high-metallicity limit and models of this phase of evolution are not yet sufficiently accurate to say which side of this limit each of the populations are on. However, the models of \citet{MG07} and \citet{KL07} both suggest the metal-intermediate population is least likely to produce carbon stars, while carbon stars may form in the metal-rich and (possibly) metal-poor populations (note that Figure \ref{MaxCOFig} assumes a solar [$\alpha$/Fe] ratio). This has already been suggested by \citet{LZS+09}, who claim the most-highly-reddened carbon stars are metal-rich on the basis of the C$_2$H$_2$ and SiC features in their infrared spectra.

There are two important reasons why we now believe that the bulk population becomes carbon-rich, despite the above results. Firstly, we do not see any significant radial segregation of the carbon stars compared to the rest of the AGB population (Figure \ref{AGBFig}), of which they represent a significant fraction. Due to the strong metallicity gradient in the galaxy \citep{SDM+07}, such a radial segregation would occur if carbon stars only arise from a fraction of the population.

Secondly, the stellar death rate correlates with the (carbon-rich) planetary nebula formation rate. Three associated planetary nebulae are known within the area we define as the Sgr dSph's main body: Hen 2-436, Wray 16-423 and StWr 2-21 \citep{ZGW+06,KZG+08,OMR+11}. All have well-defined metallicities, and are known to belong to the metal-intermediate or metal-rich populations. Crucially, all are carbon-rich. The oldest, StWr 2-21, has a dynamical age of 5200 years. This is consistent (within errors) with every AGB star in the bulk population becoming a carbon-rich planetary nebula, leaving a $\approx$0.61 M$_\odot$ white dwarf remnant. It is inconsistent with solely the much smaller, metal-rich populations producing planetary nebulae.

On balance, it seems highly likely that at least the majority of stars in the bulk population of the Sgr dSph become carbon rich and end their nuclear-burning lives by producing a planetary nebula. We cannot currently comment on the eventual fate of the metal-rich and metal-poor populations.

\subsection{The S stars}
\label{SStarSection}

At low metallicity, less carbon needs dredged up to increase C/O to above unity. The S-star transition should therefore occur in a shorter period of time and thus few S stars are expected. However, the wavelengths of the dominant ZrO bands in these spectra are very close to those of the TiO bands (Figure \ref{SpecFig}), making visual identification of S stars possible only when the C/O ratio is very high (C/O $\gtrsim$ 0.9, based on a comparison to the models presented in \citet{vENP+11}).

One of our S stars (\#424; $K_{\rm s}$ = 11.34 mag) is $\approx$0.7 mag below the RGB tip (Figure \ref{TypeFig}). This gives it a luminosity of $\sim$1000 L$_\odot$, placing it below the start of the TP-AGB, before significant dredge-up of carbon-enhanced material has occurred. Two possibilities may explain its existence. It may be an extrinsic S star: a secondary star in a binary system which has undergone mass transfer from a carbon-rich primary, but not quite enough to make it carbon-rich. However, the location of a carbon star at a similar magnitude, and relative lack of S or carbon stars at fainter magnitudes, suggests it has an intrinsic origin. If so, it is likely that both the faint S and faint C star are currently at a luminosity minimum, having recently gone through a thermal pulse. Carbon stars like these are seen in other environments, including the disrupted dwarf galaxy core, $\omega$ Centauri \citep{vLvLS+07}. While it is not clear whether the stars in $\omega$ Centauri are intrinsic or extrinsic either, they cluster near the RGB tip and are thus more likely to be intrinsic\footnote{Note, however, that decreasing ZrO band strength with temperature makes hotter S stars more difficult to identify \protect{\citep{ZBM+04}}.}. The remaining four S stars (\#361, \#909, \#942 and \#1054) are almost certainly intrinsic S stars.

The brightest S star, (\#361; $K_{\rm s}$ = 9.28 mag) is above the obvious AGB tip ($K_{\rm s} = 9.50$ mag). Similarly to \#424, there are three possibilities to explain this. Firstly, this may be a variable star caught in a luminous part of its pulsation cycle (variations of $\Delta K_{\rm s} \sim 2$ mag are not unusual; cf.\ \citet{RMSB03}). Secondly, the star may be currently undergoing a thermal pulse and thus be abnormally bright (though this phase is very short in comparison to the thermal pulse cycle). Thirdly, it may be a more-massive star in the metal-rich population, which survives to significantly higher luminosities on the AGB.

If every AGB star in the bulk population does indeed become carbon rich, we can use the evolutionary rates in Section \ref{EvolRateSect} to determine the speed of carbon enrichment in that population. There are 700 objects brighter than the faintest S star (\#424) and redward of our colour cut within the Sgr dSph core region. Of these, 114 are confirmed Sgr dSph members and 4 non-members. We can therefore expect of order $5 / 114 \times 114 / 118 \times 700 \approx 30$ ($\pm$ 13) S stars in the Sgr dSph core. If one star evolves past a certain point every 3500--5700 years, the transition from C/O $\approx$ 0.9 to C/O $\approx$ 1.0 should take of order 60--250 kyr. This implies that a single dredge-up event (which happens every thermal pulse, or $\sim$100 kyr) typically enriches the stellar surface with between $\sim$4\%--17\% of the star's initial carbon abundance. This value may increase (slightly) further if star \#424 is an extrinsic S star. Such values are not unexpected for low-mass, low-metallicity stars (e.g.\ \citealt{MG07}). If the amount of carbon enrichment exceeds $\sim$10\% per thermal pulse, many stars may progress straight from the oxygen-rich M-star phase to the carbon-rich phase, without passing through the S star phase at all. However, if carbon enrichment per pulse is lower, stars might survive one or two thermal pulses in the S-star phase.

\subsection{The carbon-rich population}
\label{CStarSection}

\subsubsection{Carbon richness and dust production}
\label{CDustSection}

With a single exception (\#588, $K_{\rm s}$ = 13.53 mag), all the carbon stars identified in the Sgr dSph are either within one magnitude of, or above, the RGB tip. This argues that the vast majority of carbon stars in the galaxy are intrinsic carbon stars, excepting \#588. The extent of the oxygen-rich AGB to $\approx$1.2 magnitudes above the $K_{\rm s}$-band RGB tip further suggests that the transition to C/O $>$ 1 typically (but not necessarily always) occurs above the RGB tip. Unsurprisingly, this is also the region where thermal pulses are active, and the convective zone of the stellar atmosphere allows large amounts of carbon-enhanced material to be brought to the surface.

The oxygen-rich AGB tip is defined by 2MASS 18514859--3120157 at ($J-K_{\rm s}$) = 1.30 mag and $K_{\rm s} = 9.50$ mag. Using the SED-fitting method of \citet{MvLD+09}, incorporating the additions used in \citet{MJZ11}, we model the luminosity of this star using the BT-Settl models of \citet{AGL+03}, assuming [Fe/H] = --0.55 dex, $M$ = 1.1 M$_\odot$, $d$ = 26.3 kpc and $E(B-V)$ = 0.14 mag. We arrive at a luminosity of 5200 L$_\odot$ for the AGB tip, with an associated error imparted by dust-enshrouding, stellar variability, uncertain distance and incomplete photometry estimated at $\pm$500 L$_\odot$. With the exception of the aforementioned S star (\#361), all stars beyond this point appear to be highly-reddened carbon stars.

Close to the RGB tip, stars also start to diverge from the main giant branch toward much redder ($J$--$K_s$) colours (region (c) of Figure \ref{CutsFig}). This indicates the presence of dust production in these stars. We can note that all the \emph{observed} stars which lie redward of the giant branch are carbon stars, which strongly suggests that \emph{all} of the stars in region (c) are carbon stars\footnote{Changes in molecular opacity in carbon stars also lead to reddening compared to otherwise-identical oxygen-rich stars, but this accounts for, at most, a few tenths of magnitudes in $(J-K_{\rm s})$.}. This is perhaps unsurprising, as dust products of carbon stars (notably amorphous carbon) are much more opaque, especially in the near-infrared, than the dominant sources of opacity in oxygen-rich stars (silicates and possibly metallic iron). 

There also appears a population of dustless carbon stars lying within the main giant branch, but these stars are few in number compared to the dusty, reddened stars. It is possible that enhanced dust production begins to occur shortly after the star becomes carbon rich, though we cannot rule out that the carbon stars have dustless episodes (related, e.g., to their thermal pulse cycle). This pattern of a few dustless and many dusty carbon stars on the AGB is also seen in other (mostly dwarf) galaxies with populations of similar age and metallicity \citep{WMF+09,MFWM11,BSvL+11,JvLM11}.

\subsubsection{Carbon star lifetimes}

We can compute the average time stars spend in the carbon-rich phase by performing a similar analysis to that in Section \ref{SStarSection}. We note that individual carbon stars may have considerably different lifetimes than this average.

We first require the number of intrinsic carbon stars in the Sgr dSph. As we discussed in the previous section, it is likely that all the reddened stars (defined by region (c) in Figure \ref{CutsFig}) in the Sgr dSph are carbon stars. Whether region (e) also contains a carbon star population is uncertain, and we therefore have between 29--52 reddened carbon stars in the Sgr dSph core, as a temporal average. Again, we split the giant branch as before into regions (a) and (b). Above the RGB tip, region (b) contains an estimated 4--15 carbon stars in this region (Table \ref{CountsTable}). In region (a) (below the RGB tip), there are 700 2MASS objects brighter than our second-faintest\footnote{Again, we assume that \#588 is an extrinsic carbon star and thus ignore it.} carbon star (\#1044, at $K_s = 11.533$ mag). We estimate that there are 5--20 carbon stars in this region. Adding these contributions together makes between 38 and 87 intrinsic carbon stars in the Sgr dSph core in total (note that, at the present epoch, we estimate there to be between 45 and 77 such stars).

If these are mostly intrinsic AGB carbon stars, and are representative of the bulk population, then our previous evolutionary rate (3500--5700 years) yields a typical carbon star lifetime of 130--500 kyr, or approximately two to five thermal pulses. 

We note that a carbon star lifetime of 130--500 kyr is much shorter than the typical TP-AGB lifetime (530--1330 kyr; Section \ref{EvolRateSect}), therefore its carbon stars should never reach C/O ratios much higher than unity (perhaps averaging C/O$_{\rm max}$ $\sim$ 1.2--1.5, on the basis of $\Delta$ C/O $\sim$ 0.1 every $\sim$100 kyr (Section \ref{SStarSection}). The Sgr dSph bulk population is evidently close to the limit for producing carbon-rich stars (indeed, \citet{MG07} predict the bulk population should not become carbon-rich; see Figure \ref{MaxCOFig}). One may therefore find it unsurprising that carbon stars have a short lifetime, as they become carbon-rich shortly before the superwind would otherwise occur.

However, it has been previously theorised (see Section \ref{IntroSect}) that process of becoming carbon rich heralds the star's death. In this scenario, the increased opacity of carbon molecules causes the star to expand, such that more of the extended atmosphere reaches the dust-producing zone. Meanwhile carbonaceous dust, being more opaque than silicaceous dust, provides increased dust-driving efficiency within the wind. This increases the mass-loss rate to such an extent that the star's atmosphere can be dissipated on a much shorter timescale. This scenario requires that the oxygen-rich progenitor star is already close to the threshold needed to generate a superwind, which seems primarily set by its pulsation characteristics \citep{vLCO+08}. This criterion may hold for the bulk population of the Sgr dSph, and it thus provides a useful laboratory in which to test such theories.

\subsubsection{Return of enriched material to the ISM}
\label{ISMSect}

\begin{figure}
\resizebox{\hsize}{!}{\includegraphics[angle=270]{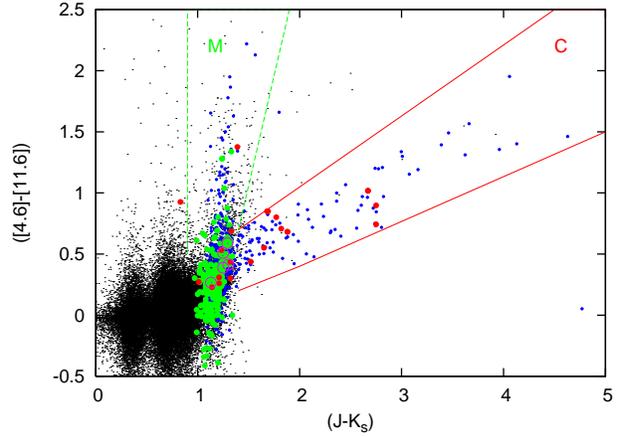}}
 \caption{2MASS--\emph{WISE} colour-colour diagram of the Sgr dSph. Black points show the entire 2MASS--\emph{WISE} cross-correlated catalogue brighter than $K_s = 11.22$ mag. Small, blue dots denote the subset of colour-selected AGB Sgr dSph stars; red dots denote observed carbon stars; green dots denote observed M-type stars. The lines mark the approximate locations of the carbon-dust-producing and silicate-dust-producing evolutionary tracks. The S stars, marked with magenta circles, all lie in the largely-dustless population near the confluence of these two tracks.}
 \label{WiseFig}
\end{figure}

We noted in Section \ref{CDustSection} that all the stars with very red ($J-K_{\rm s}$) colours appear to be carbon stars. This is unsurprising when one considers that circumstellar carbonaceous dust provides a much higher ($A_J$--$A_{Ks}$) extinction than an equivalent amount of oxygen-rich dust, and that reprocessed emission from this dust can already be significant at $K_s$-band. Conversely, the increased transparency of oxygen-rich (silicaceous) dust means that ($A_J$--$A_{Ks}$) will be low, even for significant amounts of dust. Oxygen-rich dust also tends not to re-emit effectively shortward of the Si-O bending mode near 10 $\mu$m, unless it contains a significant fraction of iron (e.g.\ \citealt{KdKW+02,VvdZH+09,MSZ+10}).

Figure \ref{WiseFig} shows a colour--colour diagram incorporating both the 2MASS and \emph{WISE} \citep{CWC+12} data. These colours have been chosen so that the heavily-reddened carbon stars, which usually display very red colours at almost all near- and mid-infrared wavelengths, lie off to the right of the diagram, as indicated by the solid red lines. Meanwhile, the oxygen-rich giants have blue ($J-K_{\rm s}$) colours due to the lack of dust opacity, but large ([4.6]--[11.6]) colours thanks to the 10-$\mu$m silicate feature, as indicated by the dashed green lines. This diagram appears to separate the different stellar types reasonably well, but is not 100\% effective. A full analysis of the relative mass loss from, and dust production by, oxygen- and carbon-rich stars is beyond the scope of this paper. However, Figure \ref{WiseFig} clearly shows that the carbon stars are not the only dust producer in the Sgr dSph: the oxygen-rich, M-type stars return significant amounts of dust as well.


\section{Conclusions}

In this work, we have performed a spectroscopic study of the nearby dwarf galaxy, the Sgr dSph. We present 1142 spectra, estimating radial velocities for 1058 and finding with high ($>$99\%) confidence that at least 592 are members. Highlights of these spectra include:
\begin{enumerate}
\item Spectra of 23 carbon stars, of which 19 are newly-determined;
\item Spectra of 5 S stars, all of which are newly-determined;
\item Spectra of $\sim$2--6 stars which may belong to the metal-poor population.
\end{enumerate}

From these spectra and associated photometry we find:
\begin{enumerate}
\item The carbon- and oxygen-rich stars separate into two distinct branches on colour--magnitude and colour--colour diagrams, leading to the estimation that there are 31 $\pm$ 14 S stars and 63 $\pm$ 12 carbon stars in the central $3^\circ \times 3^\circ$ of the Sgr dSph in total.
\item The average star in this region is separated from its nearest evolutionary neighbour by between 3500 and 5700 years of evolution. Extrapolating this to the portion of the galaxy not obscured by the Galactic Bulge gives one star per 1000--1700 years.
\item We use the observed planetary nebula frequency within this region (3) to argue that the entire bulk population appears to form carbon-rich stars, then planetary nebulae.
\item On this basis, stars spend an average of 60--250 kyr as S stars (0.9 $<$ C/O $<$ 1) and 130--500 kyr as carbon stars (C/O $>$ 1).
\item If every star becomes a carbon star, each thermal pulse should increase the C/O ratio by 0.04--0.17, and stars should typically reach a maximum C/O barely exceeding unity, though individual stars may reach higher C/O ratios.
\item Carbon stars return a significant amount of carbon-rich dust to the ISM, though it is not clear whether or not this is enough to exceed the return from oxygen-rich stars.
\item A small velocity gradient may exist, implying either tidal distortion or slow rotation around the galaxy's major axis.
\end{enumerate}

We suggest that further work is needed to determine elemental abundances of these stars, with particular emphasis on identifying their metallicities.


\section*{Acknowledgments}

We are grateful to the anonymous referee and Dr. Stefan Uttenthaler for their helpful comments on the manuscript. This publication is based on observations made with ESO Telescopes at the La Silla Paranal Observatory under programme ID 385.D-0430. It makes use of data products from the Two Micron All Sky Survey and the Wide-field Infrared Survey Explorer, which are joint projects of (2MASS) the University of Massachusetts and IPAC/CIT, funded by the NASA and the NSF, and (\emph{WISE}) the University of California, Los Angeles and JPL/CIT, funded by NASA.



\label{lastpage}

\end{document}